\documentclass[preprint,12pt]{elsarticle}

\usepackage{amssymb}
\usepackage{url}
\usepackage{amsmath}
\usepackage{xcolor}
\usepackage{soul}

\newcounter{bla}

\begin{document}

\begin{frontmatter}



\title{A multi-language auto-differentiation module and its application to a parallel particle-in-cell code on distributed computers}

\author[a]{Ji Qiang\corref{author}}
\ead{jqiang@lbl.gov}
\author[b]{Yue Hao}
\author[c]{Allen Qiang}
\author[b]{Jinyu Wan}

\cortext[author] {Corresponding author.}
\address[a]{Lawrence Berkeley National Laboratory, Berkeley, CA 94720, USA}
\address[b]{Facility for Rare Isotope Beams, Michigan State University, East Lansing, 48824, USA}
\address[c]{Stanford University, 450 Jane Stanford Way,
Stanford, CA 94305, USA}

\begin{abstract}
The auto differentiable simulation is a type of simulation that outputs of the simulation include not only the simulation result itself, but also their derivatives with respect to various input parameters. It provides an efficient method to study sensitivity of the simulation results with respect to the input parameters. Furthermore, it can be used in gradient based optimization methods for rapidly optimizing design parameters. In this paper, we present the development of a fast and transparent auto-differentiation module/class designed for easy integration into numerous simulation codes. As an application,
this auto-differentiation module is integrated into a parallel particle-in-cell code with message passing interface (MPI) on distributed memory computers. 
\end{abstract}


\end{frontmatter}

\section{Introduction}

Recent studies within the particle accelerator community have shown significant interest in auto-differentiable simulation \cite{roussel2022,roussel2023a,roussel2023b,qiang2023,cheetah,wan}.
Unlike conventional simulation methods, differentiable simulation not only yields the desired outcomes but also computes their derivatives with respect to various given parameters during the simulation process. These parameters can encompass initial charged particle beam characteristics or the accelerator lattice control and acceleration settings employed in the simulation.

The derivatives obtained from differentiable simulations offer quantitative insights into the sensitivity of the simulation results to these parameters. Such sensitivities are valuable for establishing tolerance limits for the corresponding machine parameters in accelerator design.
Moreover, these derivatives can be utilized in gradient-based optimization algorithms, such as variable metric optimization, to facilitate rapid accelerator design parameter optimization.

Automatic differentiation is a computational technique that allows for the calculation of derivatives of complex functions with respect to their variables by applying a set of fundamental function rules. This technique avoids the need for symbolic derivative representations or numerical approximations. Instead, it breaks down the function evaluation into a series of elementary operations and simple functions, using the chain rule to combine the derivatives of these basic components to determine the overall function derivatives.

It operates in two primary modes: forward mode and reverse mode. For a composite function, forward mode applies the chain rule from left to right (or inside to outside), whereas reverse mode applies it from right to left (outside to inside).

Auto-differentiation (AD) has found extensive application in the artificial intelligence/machine learning (AI/ML) community for training neural networks through gradient-based optimization \cite{ad}. Several AI/ML frameworks, including PyTorch \cite{pytorch} and TensorFlow \cite{tensorflow}, incorporate this functionality.
\begin{figure}[!htb]
\centering
\includegraphics[width=6.5 cm]{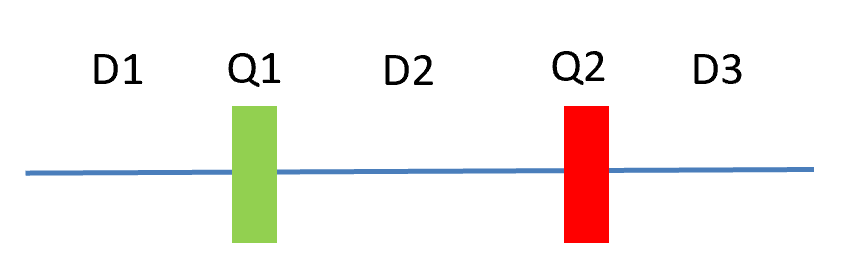}
\includegraphics[width=6.5 cm]{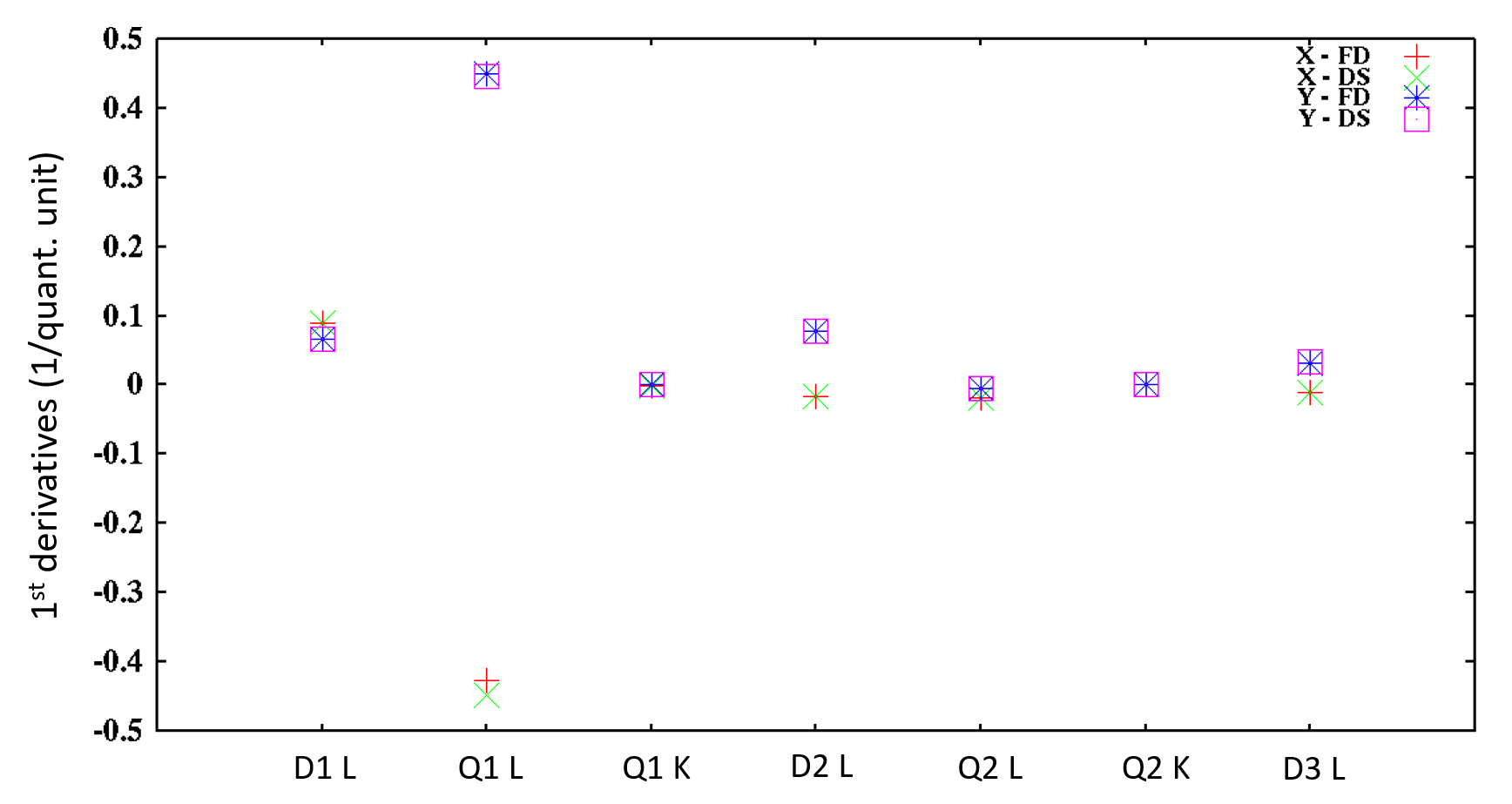}
\caption{
Layout of a FODO lattice (left) and derivatives of final horizontal and vertical emittances
with respect to seven lattice parameters from the single
differentiable self-consistent space-charge simulation and from
the finite difference approximation to the first derivative~\cite{qiang2023} (right).}
\label{fig1old}
\end{figure}  
While the forward and reverse modes of automatic differentiation can be implemented using tree structures in AI/ML, the particle accelerator community has traditionally used a method called truncated power series algebra (TPSA) to compute high-order transfer maps in beam dynamics studies \cite{berz0}. This method has also been recently applied to differentiable space-charge simulation \cite{qiang2023}.

Figure \ref{fig1old} illustrates the derivatives of the final horizontal and vertical emittances of a proton beam with respect to seven lattice parameters of a drift, focusing, drift, defocusing, drift (FODO) lattice. These derivatives were obtained from a single differentiable self-consistent space-charge simulation and compared with the first derivative approximated using finite difference methods with eight simulation results. The excellent agreement between these approaches highlights the accuracy of auto-differentiation, which yields exact results within computer precision, unlike the finite difference method that is subject to numerical truncation errors. In this example, a Fortran version of the high-order TPSA library was employed to compute the derivatives \cite{hao}. However, using the entire high-order TPSA library for first-order derivative computation is inefficient and time-consuming; in the aforementioned example, it took approximately 860 seconds on a supercomputer.

To address this efficiency challenge, we developed a fast and transparent module/class specifically for auto-differentiable simulation. This new module achieved the same result in approximately 4 seconds for the example above, representing a speedup of over 200 times compared to the standard TPSA library.

Although auto-differentiation has been utilized in space-charge studies, to the best of our knowledge this
technique has not previously been applied to an MPI-based, distributed-memory parallel particle-in-cell (PIC) code that has broad applicability across numerous fields in addition to particle accelerator physics~\cite{liewer,decyk,lyster,reynders,wang,vu,impact,impact-t,amundson,opal}.
In this work, in addition to developing the multi-language AD module/class, 
the module/class was integrated into a
two-dimensional parallel PIC code employing MPI for the 
self-consistent simulation of high-intensity beams.

The organization of the paper is as follows: 
after the Introduction, we discuss the implementation of the AD module/class in Section 2. Section 3 presents the test and application of the AD module/class to single particle tracking. In Section 4, we discuss the incorporation of the
AD module/class into a MPI-based parallel particle-in-cell code. Conclusions are drawn
in Section 5.

\section{Implementation}
This fast module adopts the TPSA method but applies it exclusively to the first derivative. This significantly simplifies the arithmetic operations and the implementation within the module. This approach is also known as the dual number implementation of forward mode auto-differentiation in the AI/ML community\cite{adwiki}.

In the TPSA method, calculating a function's derivatives with respect to its variables is transformed into evaluating the function of a vector-like AD variable according to specific algebraic rules. For the first derivative in our module, we define an AD variable $F$ where the first element is the function value, and the subsequent elements are the partial derivatives with respect to each variable of interest: 
$F=(f,f_{x_1},f_{x_2},\cdots,f_{x_n})$. Here, $f$ is the function value, and $f_{x_i}$ represents the partial derivative with respect to variable $x_i$. For an individual variable $x_i$, its corresponding vector is $X_i = (x_i,0,0,\cdots,1,\cdots)$, where $1$ is located at the $(i+1)^{th}$ element.

The addition rule for two AD variables $F$ and $G$ is defined as:
\begin{equation}
F+G = (f+g,f_{x_1}+g_{x_1},f_{x2}+g_{x_2},\cdots,f_{x_n}+g_{x_n})
\end{equation}
The multiplication rule is:
\begin{equation}
FG = (f g,gf_{x_1}+fg_{x_1},gf_{x_2}+fg_{x_2},\cdots,gf_{x_n}+fg_{x_n})
\end{equation}
The division rule is:
\begin{equation}
\frac{F}{G} = (\frac{f}{g},\frac{f_{x_1}g-g_{x_1}f}{g^2},
\frac{f_{x_2}g-g_{x_2}f}{g^2} \cdots,\frac{f_{x_n}g-g_{x_n}f}{g^2} )
\end{equation}
The function of a AD variable $h(F)$ can be represented as:
\begin{equation}
h(F) = (h(f),h_ff_{x_1},h_ff_{x_2},\cdots,h_ff_{x_n})
\end{equation}
where $h_f$ denotes the partial derivative of $h$ with respect to $f$, i.e., $\partial h/\partial f$.

\begin{figure}[!htb]
\centering
\includegraphics[width=5.5 cm]{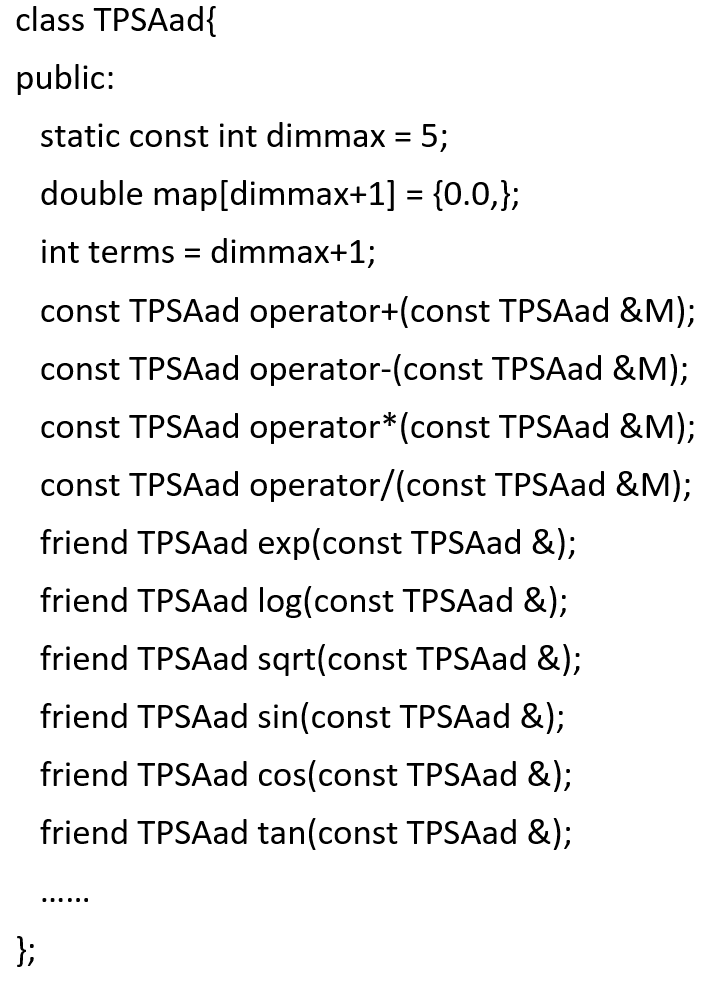}
\caption{
A section of the auto differentiation module class definition in C++.}
\label{fig3}
\end{figure}

Based on these operational rules, we can define a new data type or class for this AD vector, along with its corresponding operators and common mathematical functions. Figure \ref{fig3} shows part of the C++ implementation of this class definition. It contains three data members: one for
the number of variables to be differentiated, one for
the size of the derivative array, and the one
the derivative array storing the function value and its derivatives with respect to the specified variables.
In addition to the data members, this
class also contains a number of overloaded arithmetic operator
functions and common mathematical functions such as
polynomial, exponential, logarithm, and trigonometric functions.
A multi-language implementation of the above AD class using modern Fortran,
C++, Java, Python, and Julia is given in reference~\cite{mad}.

\section{Test and application to single particle tracking}
\begin{figure}[!htb]
\centering
\includegraphics[width=8.5 cm]{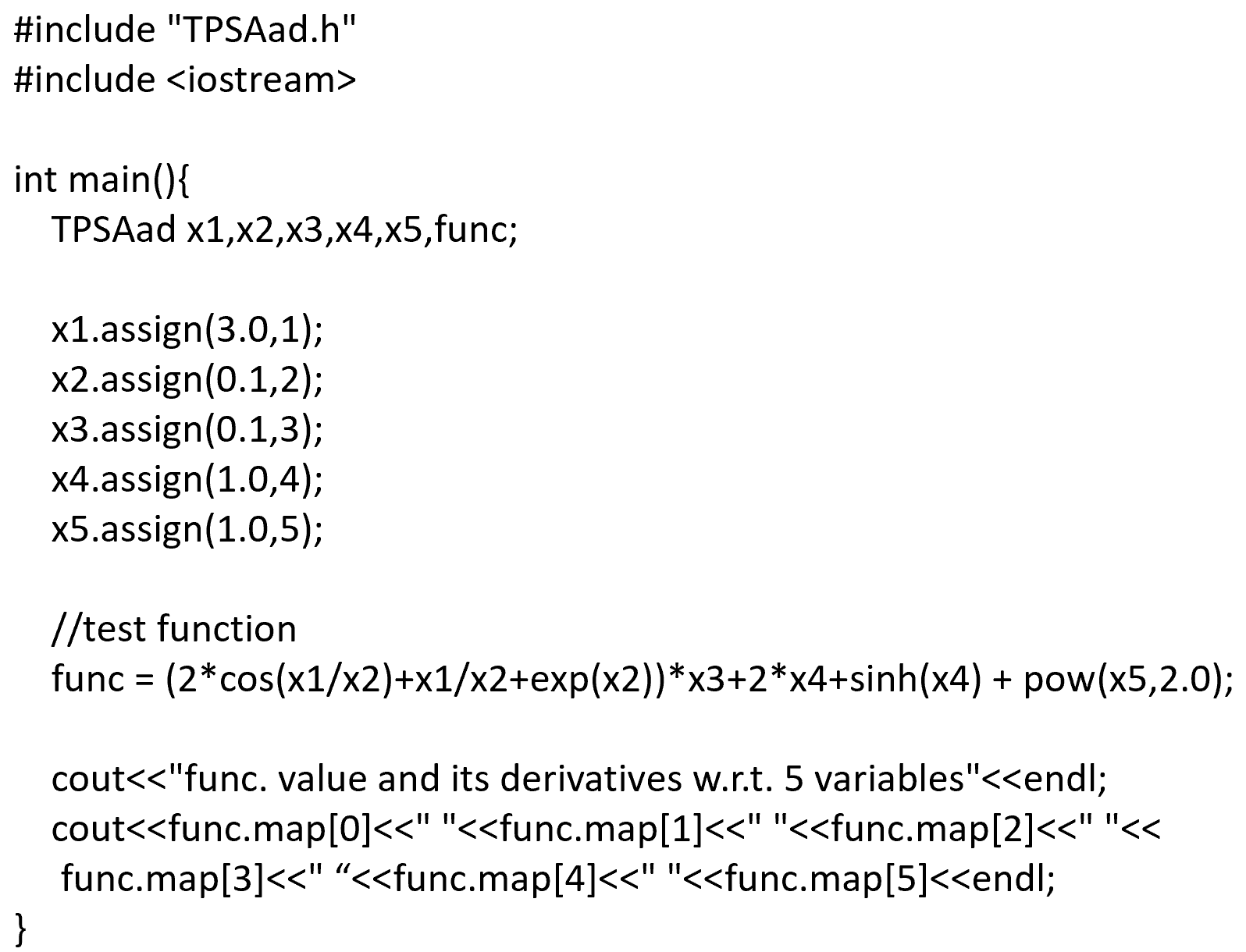}
\caption{
A C++ code showing the usage of the AD class module.}
\label{fig4}
\end{figure}  
In the following, we will present several application examples for the
above fast auto differentiation module.
Figure~\ref{fig4} shows a test example to compute a multi-variable function $f(x_1,x_2,x_3,x_4,x_5) = x_3(2cos(x_1/x_2)+x_1/x_2+e^{x_2})+2x_4+sinh(x_4)+x_5^2$ and
its derivative with respect to these five variables at variable space $(3,0.1,0.1,1,1)$ location.
The function value and its derivatives are also printed out.
In order to use the AD module/class, we need to declare the
function name and the five variable names as objects of the AD class.
The expression of the function form stays the same.
This shows that to use the AD module in a computer program, we can
keep most part the program except replacing the original data type of the function and the variables that one would like to compute derivatives
with this new data type. 

In the second example, we examined how the Twiss parameters $(\beta_x, \alpha_x, \beta_y, \alpha_y)$ change after passing through the FODO channel illustrated in Fig.~\ref{fig1old}. 
These parameters describe the shape and orientation of a particle beam's phase space ellipse and 
will evolve according
to the single particle magnetic optics inside the particle accelerator. The evolution of these parameters from starting location $0$ to location $s$ is given by~\cite{wiedemann}:
\begin{equation}
\left(\begin{array}{c}
\beta(s) \\
\alpha(s) \\
\gamma(s) 
\end{array} \right) =   \left( \begin{array}{ccc}
		m^2_{11} & -2m_{11}m_{12} & m^2_{12} \\
		-m_{11}m_{21} & m_{11}m_{22}+m_{12}m_{21} & -m_{12}m_{22} \\
        m^2_{21} & -2m_{21}m_{22} & m^2_{22} \\
	\end{array} \right)
\left(\begin{array}{c}
\beta(0) \\
\alpha(0) \\
\gamma(0) 
\end{array} \right)
\end{equation}
where $m_{ij}$ are the elements of linear transfer matrix
from location $0$ to location $s$, and $\beta \gamma = 1+\alpha^2$.
The linear transfer matrix for a quadrupole magnet is given in
Eq.~\ref{quadtr} of the next section.

We specifically assessed the sensitivity of these parameters with respect to seven key FODO lattice parameters: the lengths of the first drift space (i.e. no magnetic element) (D1 L), the first quadrupole (Q1 L), the focusing strength of the first quadrupole (Q1 k), the length of the second drift space (D2 L), the length of the second quadrupole (Q2 L), the focusing strength of the second quadrupole (Q2 k), and the length of the third drift space (D3 L). For this analysis, the normalized quadrupole strengths were set to $22 \, m^{-2}$ and $-22 \, m^{-2}$, respectively. We defined the relative change of a Twiss parameter as $\frac{xdT}{Tdx}$, where $T$ represents $\beta_x, \alpha_x, \beta_y,$ or $\alpha_y$, and $x$ is the lattice parameter being varied.
The results obtained from the auto-differentiable simulation showed excellent agreement with those calculated using the central finite difference approximation. These findings indicate that the beta function values exhibit the highest sensitivity to the length and strength of the quadrupoles.
\begin{figure}[!htb]
\centering
\includegraphics[width=5 cm]{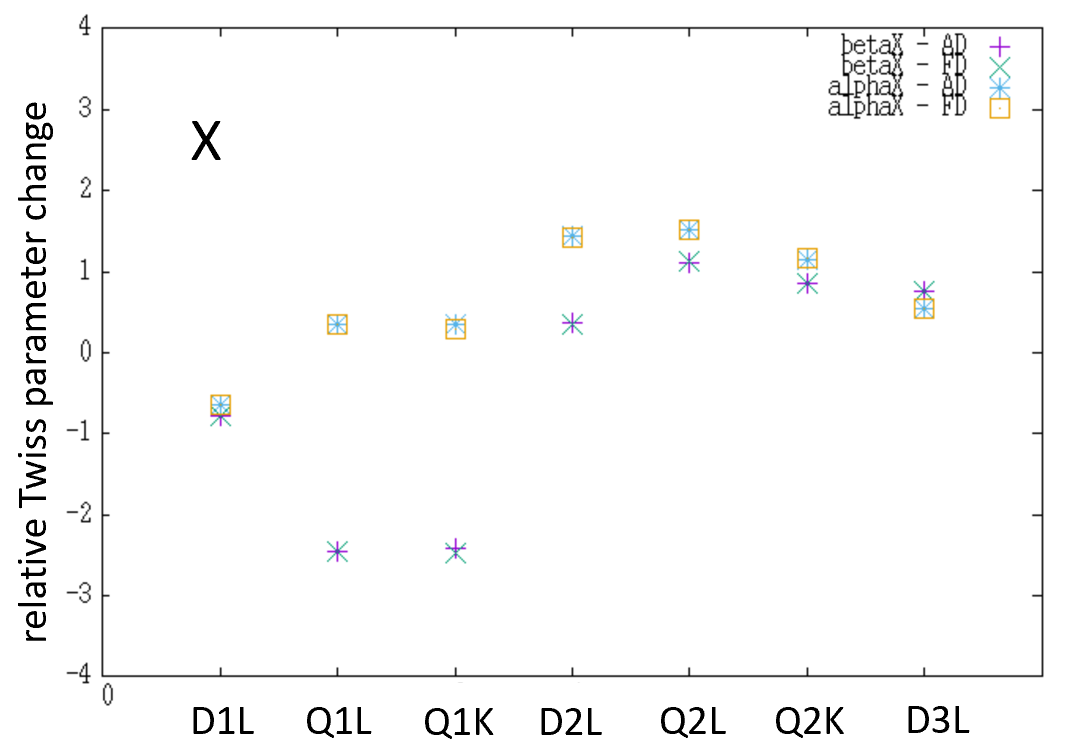}
\includegraphics[width=5 cm]{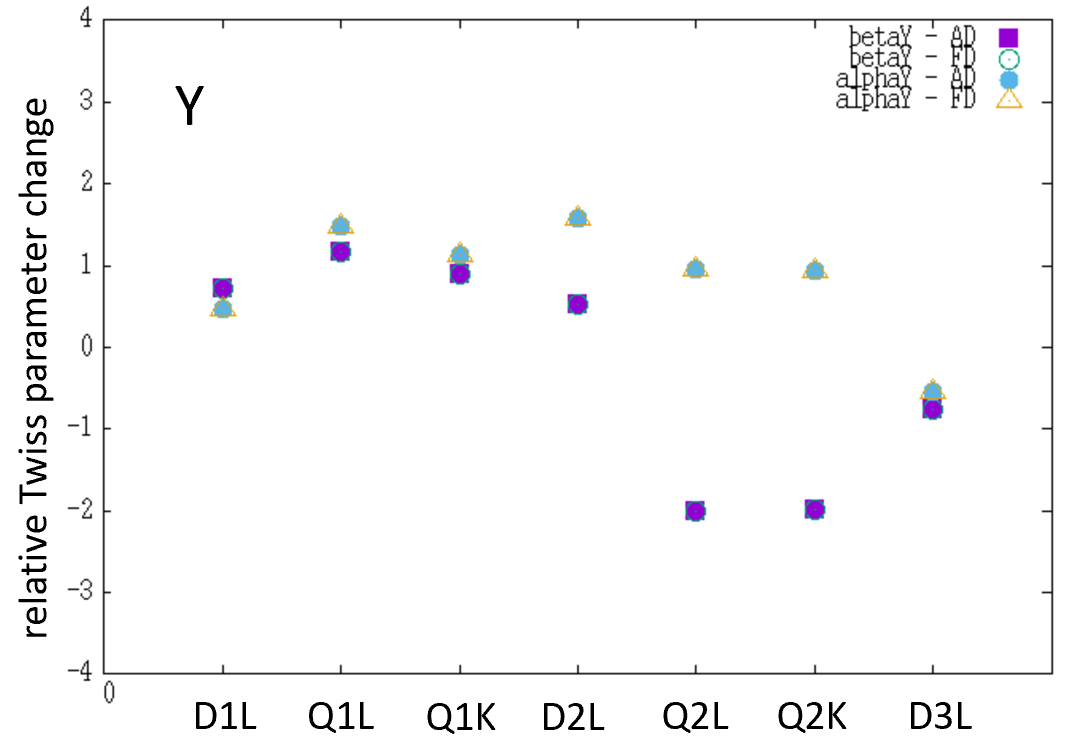}
\caption{
Relative Twiss parameter changes with respect to 7 lattice
parameters from a single auto differentiable simulation and from
central finite difference approximation with 14 simulations.}
\label{fig5}
\end{figure}  

\begin{figure}[!htb]
\centering
\includegraphics[width=8. cm]{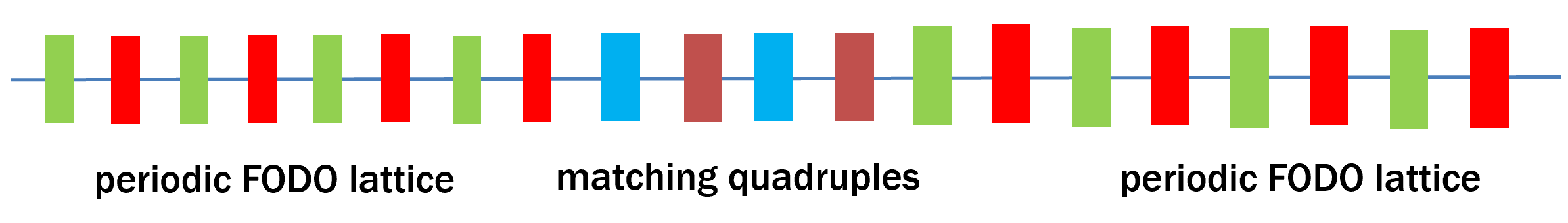}
\caption{
Schematic plot of two periodic FODO lattices and a section
of matching section between them.}
\label{fig6}
\end{figure}

The derivatives from the differentiable simulation can be used with a gradient-based parameter optimizer 
for accelerator lattice control parameter optimization.
In the third example, we integrated the differentiable simulation module
with a gradient optimizer to attain the quadrupole strengths
inside a matching section between two periodic FODO lattices.
A schematic plot of the matching section lattice and the periodic 
FODO lattices is shown in Fig.~\ref{fig6}. It consists of a periodic 
FODO lattice, a quadrupole matching section, and another periodic FODO
lattice. The four quadrupoles in 
the matching section were used
to match the given Twiss parameters 
at the entrance to the second periodic FODO lattice. 
The gradient based variable metric optimization method by Broyden-Fletcher-Goldfarb-Shanno (BFGS)~\cite{nr} was used
to minimize the objective function that is defined as follows:
\begin{eqnarray}
	f({\bf k}) & = & \frac{(\beta_x({\bf k})-\beta_{xt})^2}{\beta_{xt}^2} + 
(\alpha_x({\bf k})-\alpha_{xt})^2 + \nonumber \\
& & \frac{(\beta_y({\bf k})-\beta_{yt})^2}{\beta_{yt}^2} +
(\alpha_y({\bf k})-\alpha_{yt})^2 
\end{eqnarray}
where ${\bf k}$ is a set of control variables,
$\alpha_{xt}$, $\beta_{xt}$, $\alpha_{yt}$, and $\beta_{yt}$ are
the target Twiss parameters at the entrance to the periodic lattice, and
the $\alpha_{x}$, $\beta_{x}$, $\alpha_{y}$, and $\beta_{y}$ are the
beam Twiss parameters computed from the differentiable simulation. 
These Twiss parameters depend on the 
focusing strengths
of the quadrupoles inside the matching section. These strengths
are control variables in the above objective function.
Using the above differentiable simulation module,
the first derivatives of the objective function with respect to the
four control variables were obtained in addition to the objective
function value. These derivatives are used to construct 
an approximation to the inverse Hessian matrix in local quadratic function that guides the search for the
minimum solution.
\begin{figure}[!htb]
\centering
\includegraphics[width=7. cm]{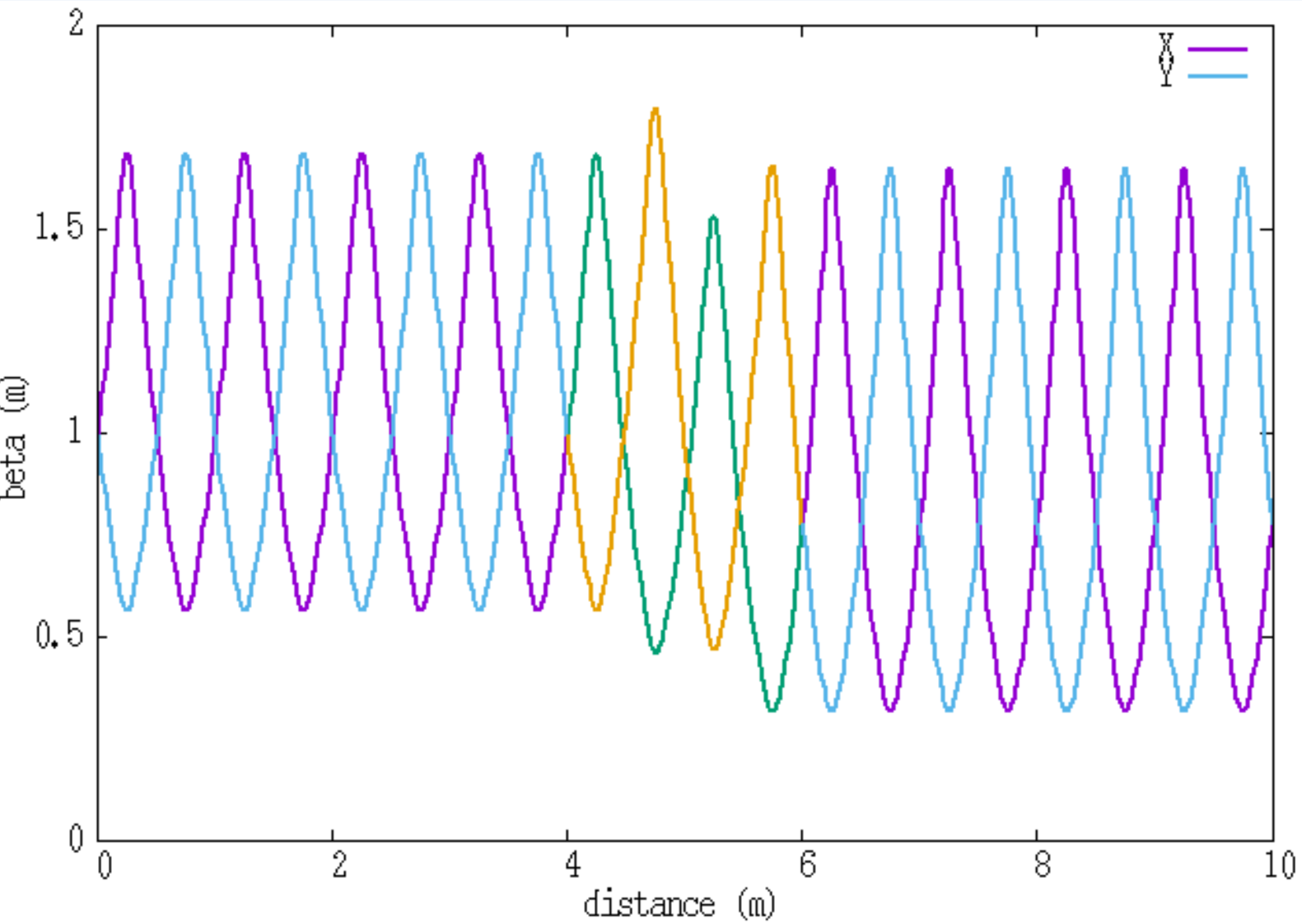}
\caption{
Evolution of Twiss beta function value through
the above lattice after optimal matching.}
\label{fig7}
\end{figure}  
Figure~\ref{fig7} shows the evolution of the transverse beta functions through the FODO lattice. As can be seen in the figure, the beta function transitions smoothly from one periodic FODO lattice to another, achieved by optimizing the strengths of the four matching quadrupoles.

The auto-differentiation incorporated into the single particle Twiss parameter tracking enables the computation of both
the objective function value and its derivatives with respect to those control parameters. Such gradient
information typically improves the speed of convergence to optimal during the search process.
\begin{figure}[!htb]
\centering
\includegraphics[width=7. cm]{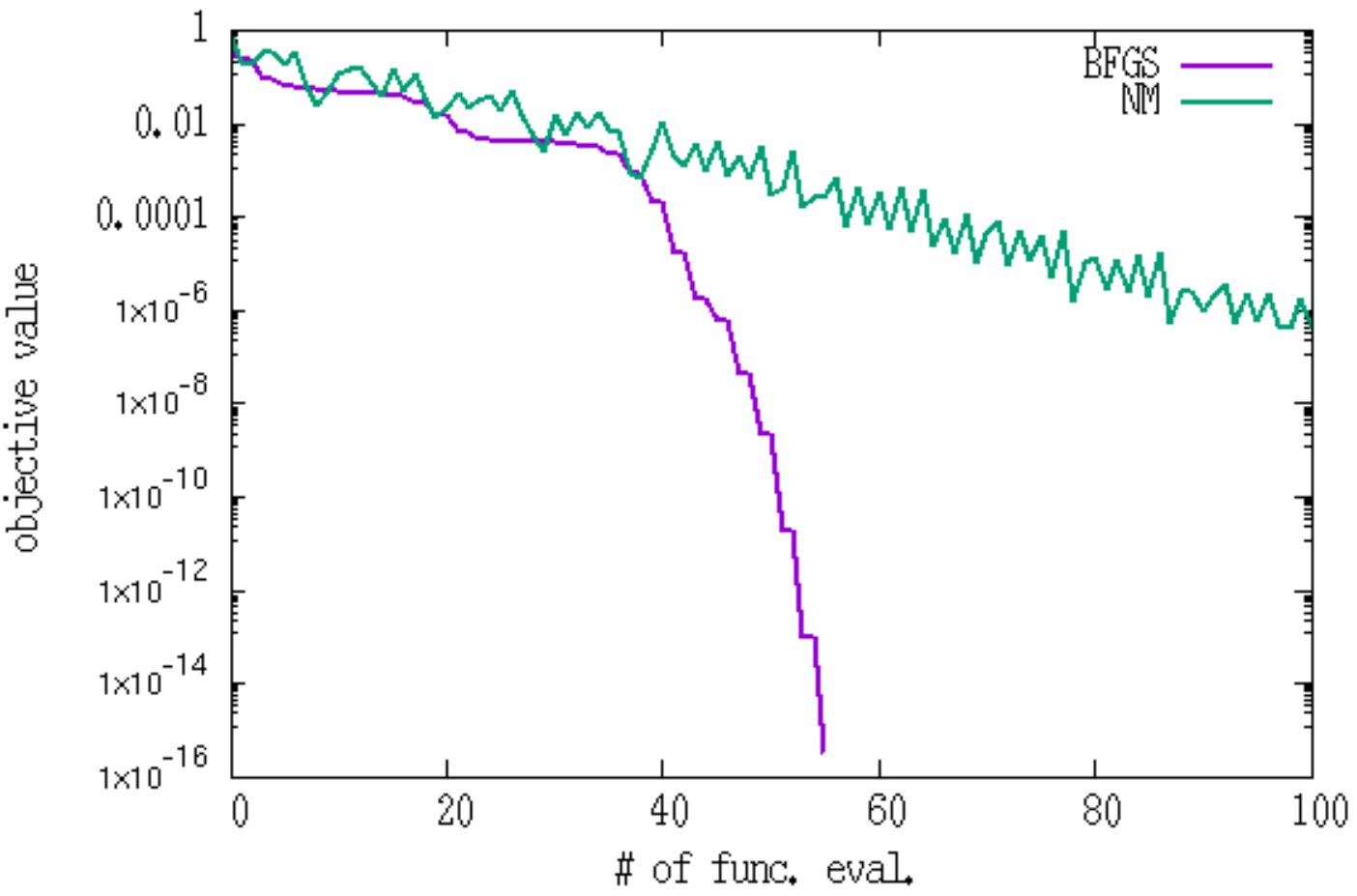}
\caption{
The objective value as a function of the number of objective function
evaluation using the auto-differentiation enabled gradient based BFGS 
method (magenta) and the derivative free simplex NM method (green).}
\label{figoptcmp}
\end{figure}  
Fig.~\ref{figoptcmp} shows the number of objective function evaluations used in the above
matching optimization example from the gradient based BFGS method and from a derivative free simplex method by Nelder-Mead (NM)~\cite{nr}.
The gradient based BFGS optimizer enabled by the auto-differentiation shows much faster convergency
compared with the derivative free simplex method.

We have implemented the above fast auto differentiation module using five programming languages: Fortran, C++, Java, Python, and Julia~\cite{mad}. To evaluate the performance of these different implementations, we measured the computational time required for a single particle to track through 1000 turns of a lattice composed of 100 periods of FODO cells.
\begin{figure}[!htb]
\centering
\includegraphics[width=7 cm]{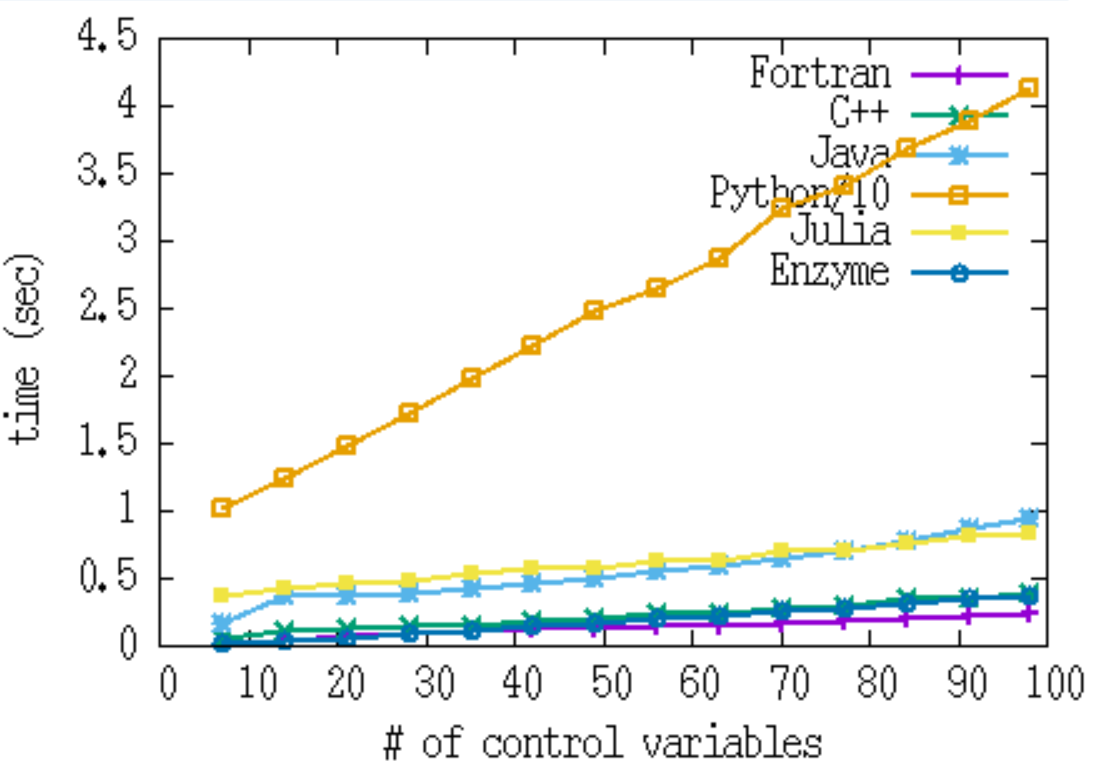}
\caption{
Computing time as a function of number of control variables with Fortran, C++, Java, Python, Julia, and Enzyme
library implementation.}
\label{fig8}
\end{figure}  
Figure~\ref{fig8} illustrates the computing time as a function of the number of control variables for the Fortran, C++, Java, Python, and Julia implementations. For comparison, we also included the computing time obtained using the well-known Enzyme package~\cite{enzyme}.
As the figure shows, the Fortran implementation exhibited the best performance, closely followed by the C++ implementation, which demonstrated comparable performance to the Enzyme library. This is perhaps expected, given Fortran's highly optimized nature for scientific computing.
The Python implementation showed the slowest performance, likely due to its interpreted nature, which necessitates runtime translation.

\section{Application to a parallel particle-in-cell code}
As another application, the fast AD module/class was integrated into a two-dimensional MPI-based parallel particle-in-cell (PIC) code that simulates the self-consistent evolution of charged particles, including space-charge effects, in particle accelerators~\cite{qiang2018}.

The charged particle dynamics is governed by 
Hamilton's equations as:
\begin{eqnarray}
	\frac{d {\bf r}}{d s} & = & \frac{\partial H}{\partial {\bf p}} \\
	\frac{d {\bf p}}{d s} & = & -\frac{\partial H}{\partial {\bf r}} 
\end{eqnarray}
where $H({\bf r}, {\bf p};s)$ denotes the Hamiltonian of the particle using distance $s$ as an independent variable, ${\bf r}=(x,y)$ denotes 
spatial coordinates of the particle, and ${\bf p}=(p_{x}/p_0,p_{y}/p_0)$ the canonical momentum coordinates of the particle normalized by the reference particle momentum $p_0$ without acceleration.
Let $\zeta$ denote a 4-vector of coordinates,
the above Hamilton's equation can be rewritten as:
\begin{eqnarray}
	\frac{d \zeta}{d s} & = & -[H, \zeta] 
\end{eqnarray}
where [\ ,\ ] is the Poisson bracket. A formal solution for above equation
after a single step $\tau$ can be written as:
\begin{eqnarray}
	\zeta (\tau) & = & \exp(-\tau(:H:)) \zeta(0)
\end{eqnarray}
Here, we have defined a differential operator $:H:$ as $: H : g = [H, \ g]$, 
for arbitrary function $g$. 
For a Hamiltonian that can be written as a sum of two terms $H =  H_1 + H_2$, an approximate
solution to above formal solution can be written as
\begin{eqnarray}
	\zeta (\tau) & = & \exp(-\tau(:H_1:+:H_2:)) \zeta(0) \nonumber \\
  & = & \exp(-\frac{1}{2}\tau :H_1:)\exp(-\tau:H_2:) \exp(-\frac{1}{2}\tau:H_1:) \zeta(0) + O(\tau^3)
  \label{ham5}
\end{eqnarray}
Let $\exp(-\frac{1}{2}\tau :H_1:)$ define a transfer map ${\mathcal M}_1$ and
$\exp(-\tau:H_2:)$ a transfer map ${\mathcal M}_2$, 
for a single step, the above splitting results in a second order numerical integrator
for the original Hamilton's equation as:
\begin{eqnarray}
	\zeta (\tau) & = & {\mathcal M}(\tau) \zeta(0) \nonumber \\
    & = & {\mathcal M}_1(\tau/2) {\mathcal M}_2(\tau) {\mathcal M}_1(\tau/2) \zeta(0)
	+ O(\tau^3)
	\label{map}
\end{eqnarray}

For the Hamiltonian in Eq.~\ref{ham5}, one can choose $H_1$ as:
\begin{eqnarray}
	H_1 & = &  {\bf p}^2/2 + q \psi({\bf r})
\end{eqnarray}
where $\psi$ denotes potential corresponding to external magnets.

For the external focusing with quadrupole magnets, the 
single step transfer map ${\mathcal M}_1$ in the focusing plane can be written as:
\begin{equation}
	{\mathcal M}_1(\tau)   =   \left( \begin{array}{cc}
		\cos(\sqrt{k}\tau) & \frac{1}{\sqrt{k}}\sin(\sqrt{k}\tau) \\
		-\sqrt{k}\sin(\sqrt{k}\tau) & \cos(\sqrt{k}\tau)
	\end{array} \right)
    \label{quadtr}
\end{equation}
and in the defocusing plane as:
\begin{equation}
	{\mathcal M}_1(\tau)   =   \left( \begin{array}{cc}
		\cosh(\sqrt{k}\tau) & \frac{1}{\sqrt{k}}\sinh(\sqrt{k}\tau) \\
		-\sqrt{k}\sinh(\sqrt{k}\tau) & \cosh(\sqrt{k}\tau)
	\end{array} \right)
\end{equation}
where $k$ is the normalized focusing strength $k=q g/p_0$ and $g$ is the 
magnetic field gradient.

For the space-charge Hamiltonian $H_2({\bf r})$, the single
step transfer map ${\mathcal M}_2$ can be written as:
\begin{eqnarray}
	{\bf r}(\tau) & = & {\bf r}(0) \\
	{\bf p}(\tau) & = & {\bf p}(0) - \frac{\partial H_2({\bf r})}{\partial {\bf r}} \tau
	\label{map2}
\end{eqnarray}
where $H_2$ accounts for the contribution from
the space-charge effects given by:
\begin{eqnarray}
	H_2 & = & \frac{1}{2}K \varphi({\bf r})
\end{eqnarray}
where $K = q I/(2\pi \epsilon_0p_0 v_0^2 \gamma_0^2)$, and $\varphi$ is the space-charge potential from the solution of the Poisson's equation.
In this Hamiltonian, the effects of the direct Coulomb 
electric potential and the
longitudinal vector potential are combined together.
The electric Coulomb potential $\varphi$ in the Hamiltonian 
$H_2$ is the solution of the Poisson equation:
\begin{equation}
\frac{\partial^2 \varphi}{\partial x^2} +
\frac{\partial^2 \varphi}{\partial y^2} = - \frac{\rho(x,y)}{\epsilon_0}
\label{poi2d1}
\end{equation}
where the charge density $\rho(x,y)$ can be obtained 
on a computational grid from
the macroparticles using a deposition scheme, e.g. a bilinear deposition scheme, in the particle-in-cell simulation. 

In this parallel particle-in-cell code, we assumes that the charged particles are confined within a rectangular, perfectly conducting pipe of the accelerator.
The boundary conditions for the electric potential $\varphi(x,y)$ inside the rectangular 
perfectly conducting pipe are:
\begin{eqnarray}
	\label{bc1}
\varphi(x=0,y) & = & 0  \\
\varphi(x=a,y) & = & 0  \\
\varphi(x,y=0) & = & 0  \\
\varphi(x,y=b) & = & 0  
	\label{boundary}
\end{eqnarray}
where $a$ is the horizontal width of the pipe and $b$ is the vertical width
of the pipe. 

Given the boundary conditions in Eqs.~\ref{bc1}-\ref{boundary}, the electric potential $\varphi$ and the
source term $n(x,y)$ can be approximated using two sine functions as:
\begin{eqnarray}
	\rho(x,y)  = \sum_{l=1}^{N_l}\sum_{m=1}^{N_m} \rho^{lm} \sin(\alpha_l x) \sin(\beta_m y) \\
	\varphi(x,y)  =  \sum_{l=1}^{N_l}\sum_{m=1}^{N_m} \varphi^{lm} \sin(\alpha_l x) \sin(\beta_m y) 
\end{eqnarray}
where
\begin{eqnarray}
\label{rholm}
\rho^{lm}  = \frac{4}{ab}\int_0^a\int_0^b n(x,y) \sin(\alpha_l x) \sin(\beta_m y) \ dx dy \\
\varphi^{lm}  = \frac{4}{ab}\int_0^a\int_0^b \varphi(x,y) \sin(\alpha_l x) \sin(\beta_m y) \ dx dy
\end{eqnarray}
where $\alpha_l=l\pi/a$ and $\beta_m = m \pi/b$.
By substituting above expansions into the
Poisson Eq.~\ref{poi2d1} and making use of the orthonormal condition of the sine functions,
we obtain
\begin{eqnarray}
	\varphi^{lm} & = & \frac{\rho^{lm}}{\epsilon_0\gamma_{lm}^2}
	\label{odelm}
\end{eqnarray}
where $\gamma_{lm}^2 = \alpha_l^2 + \beta_m^2$. 
From the above solution of the electric potential in
the frequency domain,
the electric space-charge potential $\varphi(x,y)$ on the computational grid
can be attained by applying the inverse sine function transformation.
Given the electric potential on the grid, the space-charge fields are computed using a central finite-difference approximation. 
These fields are then interpolated back to the individual
macroparticles based on their positions for the
momentum update in map $\mathcal{M}_2$.
\begin{figure}[!htb]
\centering
\includegraphics[width=7. cm]{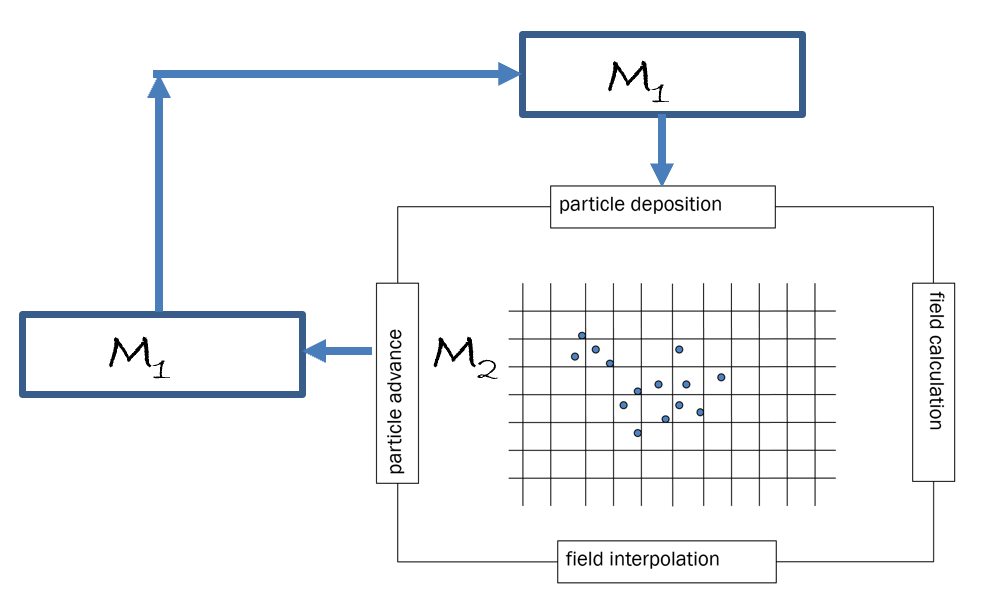}
\caption{
Schematic plot of a single step in the above particle-in-cell code.}
\label{figpic}
\end{figure}  
A single iteration step of the above particle-in-cell method is illustrated in Fig.~\ref{figpic}. 

\begin{figure}[!htb]
\centering
\includegraphics[width=5.5 cm]{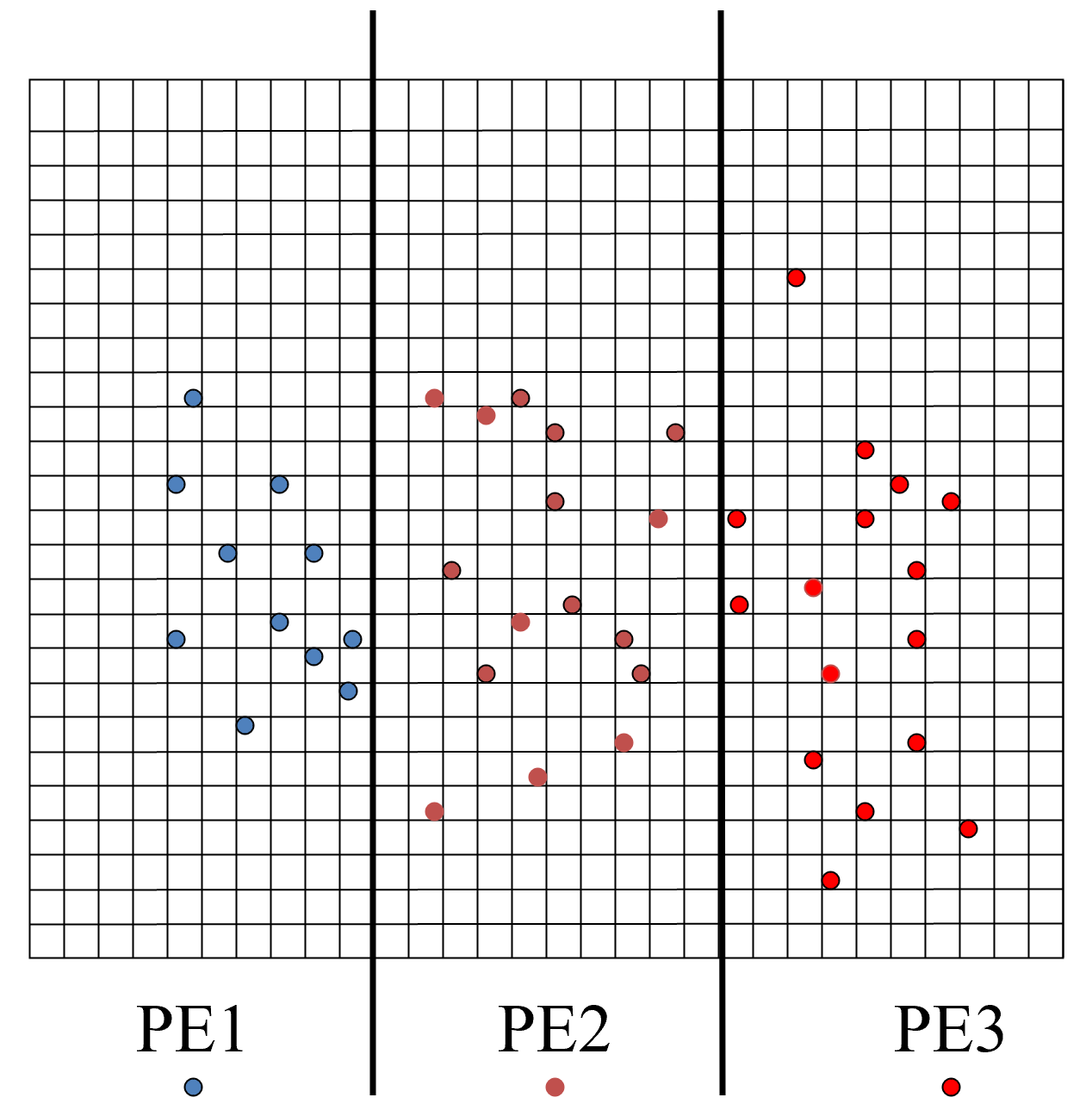}
\caption{
Schematic plot of the domain-decomposition method in the parallel particle-in-cell code.}
\label{figddeco}
\end{figure}

For parallelization, a one-dimensional (1D) domain decomposition method was adopted in this study. A schematic plot of the 1D decomposition of the computational domain is shown in Fig.~\ref{figddeco}. The solid grid lines indicate the
global computational domain grids, while the thick solid lines mark the local computational domain boundary on each processor. In this example, the
global computational domain is uniformly divided into three blocks along the $y$ dimension, each mapped to a single processor element (PE). Each processor holds $N_{ylc}=(N_y-1)/N_{proc}$ local grid points with  $N_y=31$ total grid points and $N_{proc}=3$ total processor number in this example. 
The local range of the computational domain on processor $i$ is given by $iN_{ylc}h_y\le y < (i+1)N_{ylc}h_y$, where $h_y$ is the grid
space along the $y$ dimension.
Particles with spatial positions within a processor's local computational boundary
are assigned to that processor. Since particles move around during simulation, after each step their spatial coordinates are checked against the local computational boundary. If
a particle is outside the local boundary range, this particle is sent to the neighboring processor via explicit communication for a new boundary check. 
This process is repeated until all particles reside within the local computational domain of their assigned processor.

In the PIC code, macroparticles are deposited onto a grid to obtain the charge density distribution. Depending on the shape function used in the deposition, a single macroparticle may contribute to multiple grid points. For example, in a linear deposition scheme, a macroparticle contributes to two neighboring grids in each dimension.  
In the parallel implementation, \emph{guard grids} are added outside the boundary grids to facilitate communication between processors. Here, boundary grids refer to the outermost grids within the local computational domain, while guard grids serve as temporary storage for grid quantities received from neighboring processors during local communication.  
After particles are reassigned to their local processor, charge deposition is performed locally and in parallel across all processors. In this stage, particles positioned between the local computational boundary and the boundary grid contribute not only to their local boundary grid in the linear deposition scheme but also to the boundary grid of the neighboring processor along the $y$-dimension. Neighbor-to-neighbor communication is then used to exchange the guard grid data, which are added to the charge density on the boundary grids to assemble the complete local density distribution.

Using the local charge density on the grid, the Poisson equation is solved following the spectral method described above, with a computational cost scaling as $\mathcal{O}(N \log N)$, where $N$ is the total number of grid points. In the parallel PIC code, a one-dimensional domain decomposition is employed to retain one global dimension locally for the fast sine Fourier transform (FT) in that dimension. This is followed by a data transpose and a fast sine FT in the other dimension. The transpose involves global communication to exchange data between processors, moving the local dimension into the global dimension.  
After solving the Poisson equation, the fields at grid points are computed using a finite-difference method. For boundary grid points, neighboring communication is used to retrieve potentials from the boundary grids of adjacent processors. A central finite-difference scheme is then applied on all processors to obtain the electric fields, after which neighboring communication is performed again to update boundary values.  
With the self-consistent electromagnetic fields available on the local computational grid, interpolation is used to obtain the fields on individual particles. Particle advancement is then performed on all processors in parallel for one time step. This cycle is repeated for multiple time steps until the simulation is complete.

To apply the auto-differentiation (AD) module/class, the parallel particle-in-cell code described above is rewritten using AD variables. The phase-space coordinates ${\bf r}_i$ and ${\bf p}_i$ of particle $i$, which are normally declared as double-precision variables, are instead declared as AD variables $D{\bf r}_i$ and $D{\bf p}_i$ using the module data type/class defined in Section~2.  
The map corresponding to the external quadrupole field is expressed in AD variables as  
\begin{equation}
	{\mathcal M}_1(\tau)   =   \left( \begin{array}{cc}
		\cos(\sqrt{Dk}D\tau) & \frac{1}{\sqrt{Dk}}\sin(\sqrt{Dk}D\tau) \\
		-\sqrt{Dk}\sin(\sqrt{Dk}D\tau) & \cos(\sqrt{Dk}D\tau)
	\end{array} \right)
\end{equation}
where $Dk$ is the AD variable representing the quadrupole focusing strength, and $D\tau$ is the AD variable for the step size, defined as the quadrupole length divided by the number of integration steps. A similar expression applies in the defocusing plane of the quadrupole.  
In the computation of space-charge fields, the charge density $\rho$ is replaced by the AD variable $D\rho$, and the electric potential $\varphi$ by $D\varphi$.

In the parallel implementation, MPI is used to move macroparticles among processors and to exchange boundary grid information between neighboring processors. To incorporate AD capability into the parallel PIC code, the particle phase-space coordinates and their derivatives, as defined in the AD class, are transmitted together with the original coordinates. Similarly, the density, potential, and field quantities on the boundary grids, along with their derivatives, are exchanged during MPI communication.  
The Poisson equation is solved not only for the charge density on the computational grid but also for the derivatives stored in the data member array of the $D\rho$ object.

The charged particle beam initial distribution parameters,
beam energy, and current can also be 
written using AD variables if needed.
The final beam properties such as emittances are defined
using the AD variables. 
The horizontal emittance $D\epsilon_x$ is given as
\begin{eqnarray}
	D\epsilon_x & = & \sqrt{D<x^2>D<p_x^2>-(D<xp_x>)^2}
\end{eqnarray}
where
\begin{eqnarray}
	D<x^2> & = & \frac{1}{N_p}\sum_{i=1}^{N_p} (Dx_i)^2 \\
	D<{p_x}^2> & = &  \frac{1}{N_p}\sum_{i=1}^{N_p} (Dp_{xi})^2 \\
D<{xp_x}> & = &  \frac{1}{N_p}\sum_{i=1}^{N_p} Dx_i Dp_{xi} 
\end{eqnarray}
The vertical emittance has a similar expression with $x$ replaced by $y$.

Using the same example described in the preceding section, we tested the sensitivity of the final emittances after one FODO lattice period with respect to seven lattice parameters, employing the differentiable parallel PIC simulation with a $100$~A proton beam current at $1$~GeV kinetic energy.  
\begin{figure}[!htb]
\centering
\includegraphics[width=7.0cm]{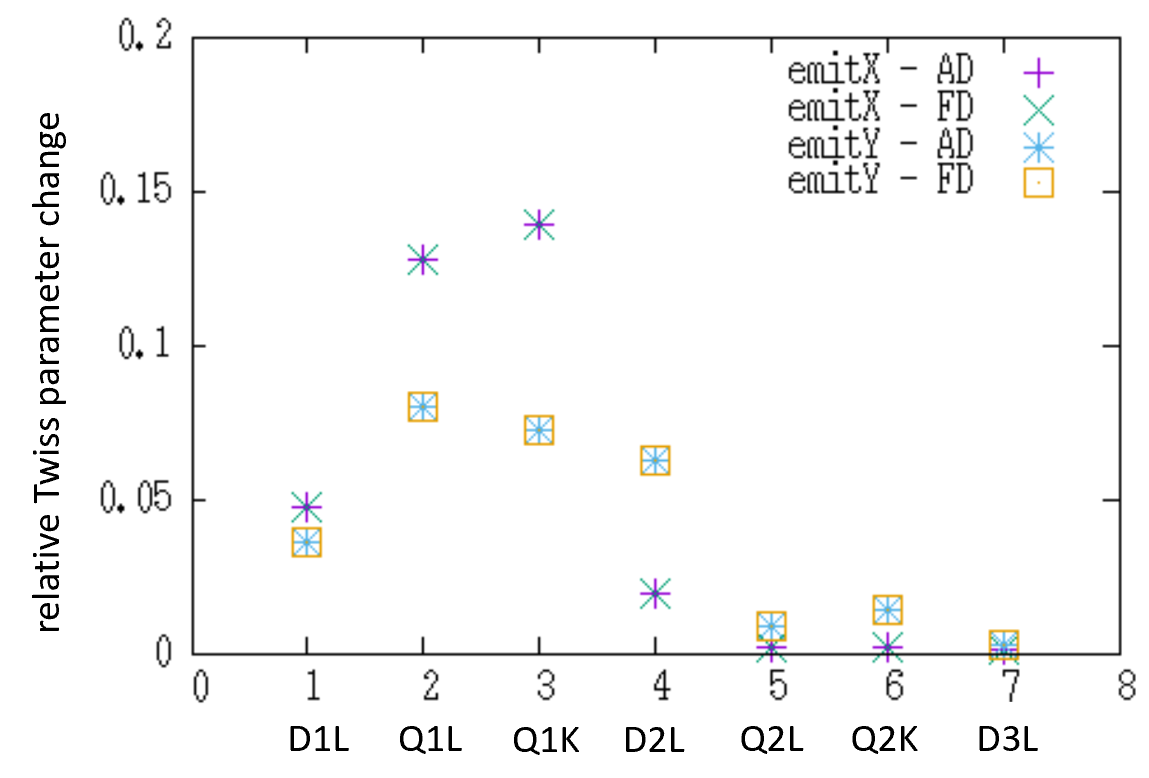}
\caption{
Relative emittance changes with respect to seven lattice parameters, obtained from a single auto-differentiable parallel PIC simulation and from the central finite-difference approximation using 14 simulations.}
\label{figemt}
\end{figure}  
Figure~\ref{figemt} shows the sensitivity of the final horizontal ($X$) and vertical ($Y$) emittances to the first drift length, first quadrupole length, first quadrupole strength, second drift length, second quadrupole length, second quadrupole strength, and third drift length. Results from the auto-differentiable parallel PIC simulation agree closely with those from the central finite-difference approximation. The final emittances are highly sensitive to the first quadrupole length and strength, and only weakly sensitive to the second quadrupole parameters and the length of the last drift.

Using the auto-differentiable parallel PIC code, we re-optimized the matching section described in the preceding section for a $100$~A proton beam current, employing the same gradient-based BFGS optimizer.  
\begin{figure}[!htb]
\centering
\includegraphics[height=3.8cm, width=5.1cm]{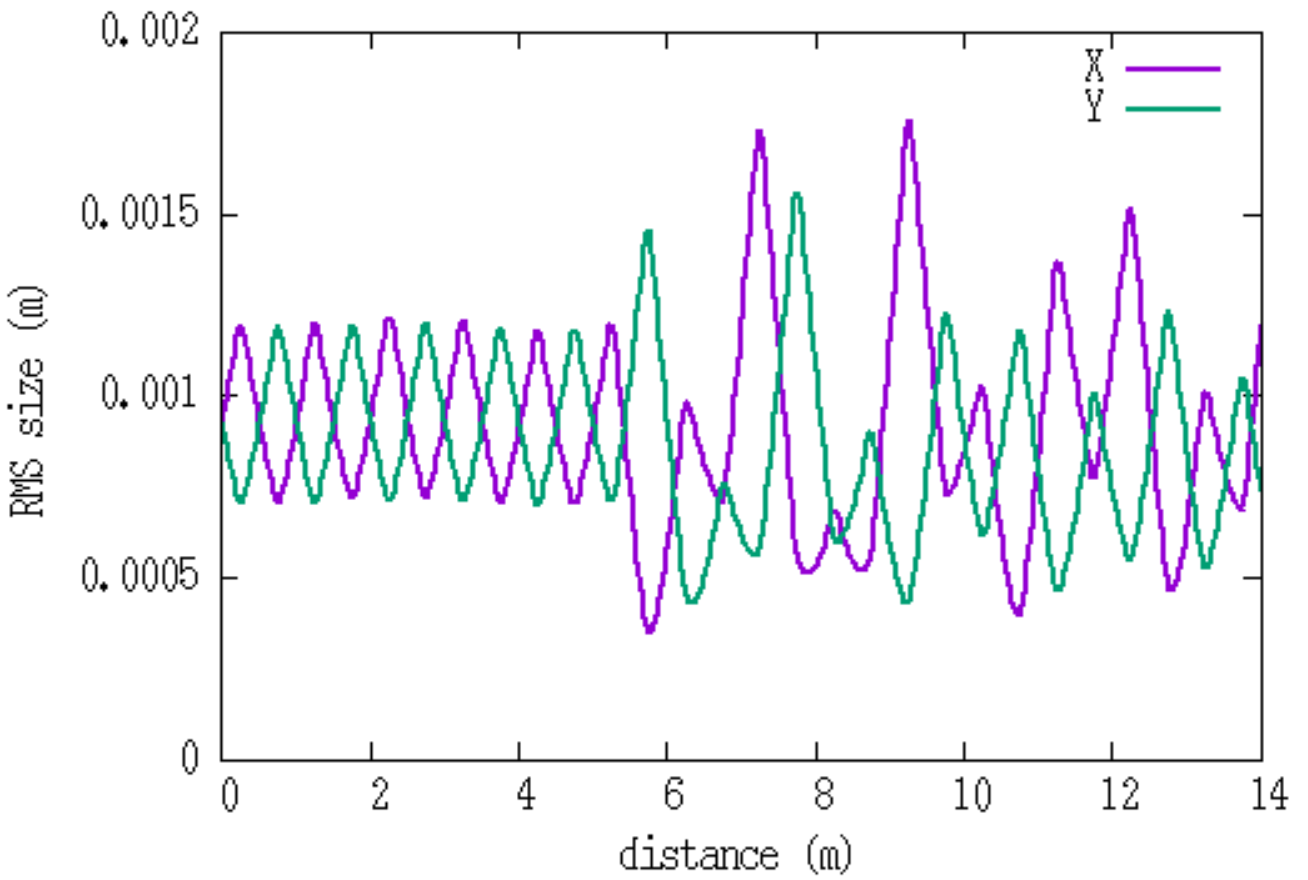}
\includegraphics[width=5.0cm]{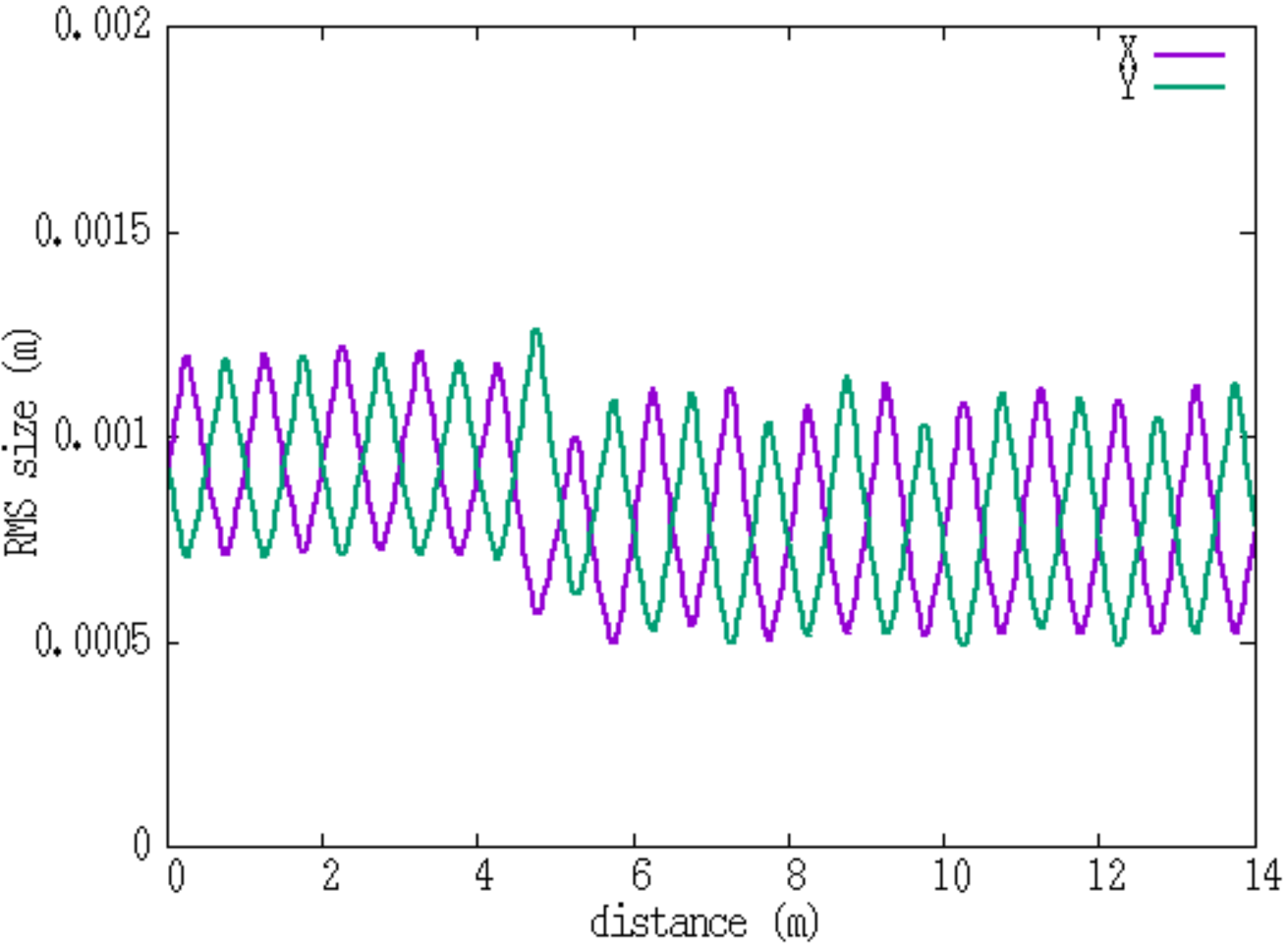}
\caption{
Evolution of the transverse root mean square (RMS) proton beam size through the FODO lattice without optimization (left) and with gradient-based optimization using the auto-differentiable parallel PIC simulation.}
\label{figrmsopt}
\end{figure}  
Figure~\ref{figrmsopt} shows the evolution of the transverse RMS beam size without and with gradient-based matching optimization from the auto-differentiable parallel PIC simulation. With the optimized quadrupole strength settings in the matching section, the transverse beam size exhibits more regular oscillations compared to the case without optimization.

\section{Conclusions}
In this study, we have developed a simple, transparent, and effective 
forward auto-differentiation (AD) module/class and implemented it using five programming
languages, Fortran, C++, Java, Python, and Julia. 
The implementation introduces a new data type/class and comprises only a few hundred lines of code, enabling straightforward integration into existing applications. Variables can be declared using this new data type without modification of the functional expressions in the original program. The class explicitly stores the derivatives of functions in a single array, which can be directly manipulated in distributed parallel computing environments using MPI.

As an illustration, 
we have shown in two representative applications: single-particle tracking and multi-particle parallel particle-in-cell simulations. 
In the parallel PIC code, MPI-based communication is achieved by transmitting derivative information stored in the AD data array to multiple processors. Incorporation of the AD class into the parallel PIC framework facilitates sensitivity analyses of final charged particle beam properties, such as emittance, with respect to accelerator machine parameters, including self-consistent space-charge effects. Furthermore, it enables efficient accelerator lattice parameter optimization through the use of gradient-based optimization algorithms.

\section*{Acknowledgement}
This work was supported by the U.S. 
Department of Energy (DOE), Office of Science, Office of 
High Energy Physics, under Contract No. DE-AC02- 
05CH11231 and DE-SC0024170. Computational resources were provided by 
the National Energy Research Scientific Computing Center 
(NERSC), which is supported by the DOE Office of 
Science under the same first contract number.





\bibliographystyle{elsarticle-num}







\end{document}